\documentclass[12pt]{article}
\usepackage{scicite}
\usepackage{epsfig}
\usepackage{times}
\newenvironment{sciabstract}{%
\begin{quote} \bf}
{\end{quote}}

\newcounter{lastnote}

\title{The Disappearance of the Red Supergiant Progenitor of Supernova 2008bk} 

\author
{Seppo Mattila,$^{1,2\ast}$ Stephen Smartt,$^{3}$ Justyn Maund,$^{4,5}$ Stefano Benetti,$^{6}$\\Mattias Ergon$^{1}$\\
\\
\normalsize{$^{1}$Department of Astronomy, Stockholm University, Oscar Klein Centre, AlbaNova,}\\
\normalsize{SE-10691 Stockholm, Sweden.}\\
\normalsize{$^{2}$Tuorla Observatory, Department of Physics \& Astronomy, University of Turku,}\\
\normalsize{V\"ais\"al\"antie 20, FI-21500 Piikki\"o, Finland.}\\
\normalsize{$^{3}$Astrophysics Research Centre, School of Mathematics and Physics,}\\
\normalsize{Queen's University Belfast, Belfast BT7 1NN, UK.}\\
\normalsize{$^{4}$Dark Cosmology Centre, Niels Bohr Institute, University of Copenhagen,}\\
\normalsize{Juliane Maries Vej, DK-2100 Copenhagen, Denmark.}\\
\normalsize{$^{5}$Sophie \& Tycho Brahe Fellow.}\\
\normalsize{$^{6}$INAF Osservatorio Astronomico di Padova, Vicolo dell'Osservatorio 5,}\\
\normalsize{I-35122 Padova, Italy.}\\
\\
\normalsize{$^\ast$E-mail: sepmat@utu.fi}}

\date{}

\begin{document} 


%\baselineskip24pt


\maketitle 

\begin{sciabstract}
Massive stars end their lives in spectacular supernova explosions. Identifying the progenitor star is
a test of stellar evolution and explosion models. Here we show that the progenitor star of the
supernova SN 2008bk has now disappeared, which provides conclusive evidence that this was the death of
a red supergiant star.
\end{sciabstract}

The progenitors of core-collapse supernovae (CCSNe) are massive bright stars,
and thus can be directly detected in nearby galaxies if high quality images
before the explosion are available.
Using this technique the progenitor stars of a total of nine CCSNe have so far
been identified; and their inferred masses all indicate that the minimum mass
threshold for a star to explode as a CCSN is 8 $\pm$ 1 M$_{\odot}$ {\it (1)}. However,
we can be absolutely confident we have correctly identified the progenitor only when the star is
confirmed to have disappeared by late time deep imaging of its site of
explosion. We have only witnessed the disappearance of the progenitors of four
CCSNe {\it (2,3,4)}, and present here such a confirmation for the nearby Type IIP
SN 2008bk. We show that the progenitor star has disappeared in four bands, which is
compelling evidence that the progenitor star's spectral type, luminosity and mass
are well constrained and that the object identified was a single star.

SN 2008bk was discovered in the nearby spiral galaxy NGC 7793 on March 2008
and was classified as a Type II-P SN. Analysis of
high-quality optical and near-infrared pre- and post-explosion images from
the European Southern Observatory (ESO) Very Large Telescope (VLT) revealed
a very red pre-explosion source coincident with the position of SN 2008bk {\it (5)}.
The colors and luminosity of the source were found consistent with it being a
M4 red supergiant star with an initial mass of 8.5 $\pm$ 1.0 M$_{\odot}$ putting it at
the low mass end of directly detected progenitors of Type II-P SNe which are
the most common type of SN explosions in the Universe, by volume. With a
detection in four different filters covering both optical and near-infrared
wavelengths this is one of the best constrained SN progenitors observed,
making its determined mass also one of the tightest constraints on the lower
mass threshold for stars to explode as CCSNe.

We have re-imaged (supporting online text) the explosion site of SN 2008bk on 2010 September 16
using the ESO Faint Object Spectrograph and Camera (EFOSC2) with $V$, $R$, and $I$
filters, and on 2010 October 29 using the SofI infrared spectrograph and imager
with $J$, $H$ and $K_{\rm s}$ filters, both on the ESO New Technology Telescope (NTT).
Using differential astrometry, we matched the pre-explosion observations
of SN 2008bk with our new images. We find no significant detection of a point source at
the SN location in the same filters ($I$, $J$, $H$, and $K_{\rm s}$) where the pre-explosion
source was detected in the VLT images obtained before the SN explosion
({\it Fig. 1}). However, SN 2008bk was still marginally detected in the $V$ and
R filters.

The disappearance of the pre-explosion source identified at the position of SN 2008bk confirms
it to be the progenitor and shows that it was a single massive star. This is compelling
verification of stellar evolution theory that predicts core-collapse when massive single
stars are red supergiants. The two best constrained and confirmed progenitors of type II-P SNe
are for SN 2003gd ({\it 6,3)} and SN 2008bk {\it (5)}. These are both M0-M4 red
supergiant stars with estimated initial masses at the lower end of the
mass range considered theoretically possible to produce a core-collapse
SN explosion.

\begin{quote}
{\bf References and Notes}

\begin{enumerate}
\item S. J. Smartt, {\it Annu.Rev.Astron.Astrophys.} {\bf 47}, 63 (2009).
\item R. Gilmozzi et al., {\it Nature} {\bf 328}, 318 (1987).
\item J. R. Maund. \& S. J. Smartt, {\it Science} {\bf 324}, 486 (2009).
\item A. Gal-Yam, D. Leonard, {\it Nature} {\bf 458}, 865 (2009).
\item S. Mattila et al., {\it Astrophys. J.} {\bf 688L}, 91 (2008).
\item S. J. Smartt et al., {\it Science} {\bf 303}, 499 (2004).
\item We thank the members of the ESO Large Programme 'Supernova Variety and Nucleosynthesis Yields'
collaboration (http://graspa.oapd.inaf.it/). We acknowledge funding from the EURYI
scheme (SJS) and the Academy of Finland project 8120503 (SM).
\end{enumerate}
\end{quote}

\bibliography{scibib}

\bibliographystyle{Science}

\clearpage

\epsfig{file=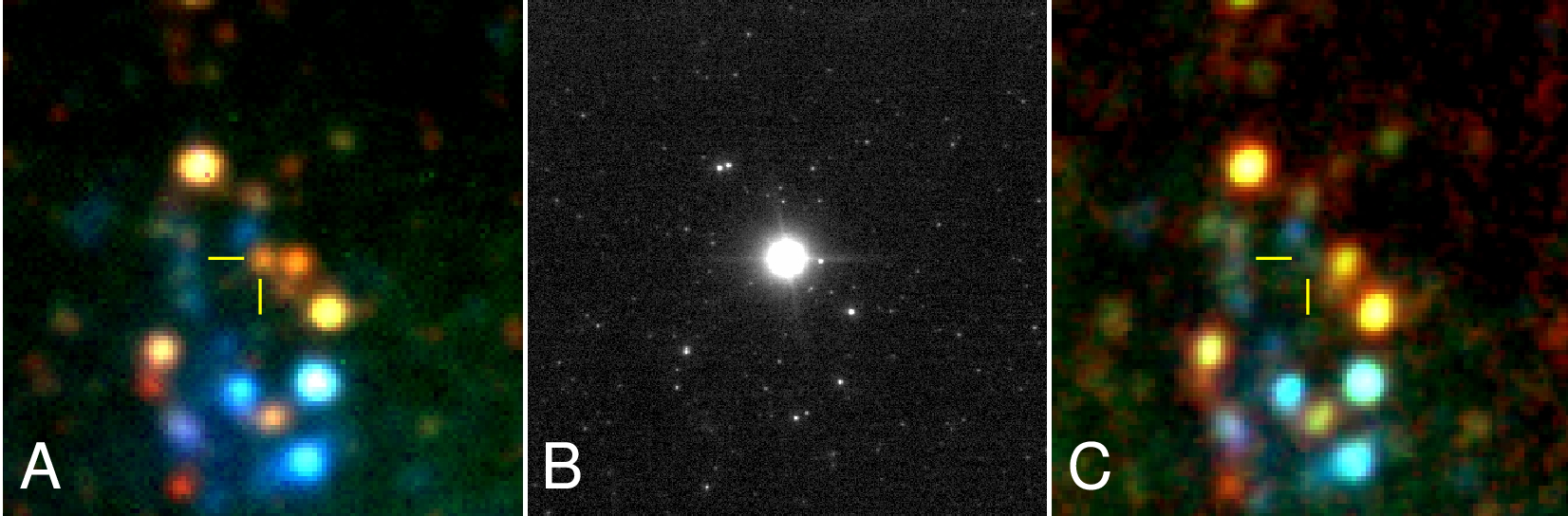,width=15cm}\\
\noindent {\bf Fig. 1.} The explosion site of SN 2008bk as observed with the ESO VLT and NTT
(supporting online text). (A) Pre-explosion color-combined ($V$, $I$ and $K_{\rm s}$-bands)
VLT image (2001 September 16 and 2005 October 17) showing a very red pre-explosion source
coincident with the SN position. (B) post-explosion $K_{\rm s}$-band adaptive optics
VLT image (2008 May 19) showing the SN near the maximum light.
(C) post-explosion late time color-combined ($V$, $I$ and $K_{\rm s}$-bands) NTT image
(2010 September 16 and October 29) observed after the SN has faded away. The red supergiant
progenitor star identified in Mattila et al. (5) is indicated in (A) but no point source
is detected at its position in the late-time image (C).

\end{document}


%\baselineskip24pt


\maketitle


\begin{sciabstract}

\end{sciabstract}

\section{Supporting Online Text}
SN 2008bk was discovered (S1) on 2008 March 25 and was spectroscopically classified
(S2) as a Type II-P supernova (SN). The site of SN 2008bk has a wealth of prediscovery
images available where a pre-explosion source was identified as a possible progenitor for the SN
(S3,S4), and in (S5) it was precisely identified as the progenitor and shown
to be a M4 red supergiant star with an initial mass of 8.5 $\pm$ 1.0
M$_{\odot}$. 

The high quality pre-explosion imaging were available from the European Southern
Observatory (ESO) Science
Archive\footnote{http://archive.eso.org/} (Table 1). The optical observations were taken with
FORS1 and the near-infrared observations with ISAAC and HAWK-I on the ESO Very Large
Telescope (VLT). To precisely identify the progenitor star we
observed SN 2008bk on 2008 May 19 using the adaptive optics imager NACO on
the VLT (program 081.D-0279, P.I.: S. Mattila). The reductions and calibration of these
pre- and post-explosion
observations are reported in (S5). The site of SN 2008bk was revisited on
2010 September 16 and October 29 when the SN was expected to have faded-away to
confirm the disappearance of the progenitor (Table 1). This time the optical
observations were taken with EFOSC2 and the near-infrared observations using
SofI on the ESO New Technology Telescope (NTT) (program 184.D-1151, P.I.: S. Benetti).
Both the optical and near-infrared images were reduced in a standard way in the
image reduction and analysis facility (IRAF).

The late time post-explosion observations were aligned in IRAF with the
pre-explosion images using geometric transformations including $x$ and $y$
shifts, and a common scale factor and rotation for $x$ and $y$ using centroid
positions of point sources common between the frames. For example, in
$K_{\rm s}$-band fifteen point sources were used yielding a root mean square
uncertainty of 35 milliarcsecond for the transformation. The comparison
between the images revealed the source identified as the progenitor of SN
2008bk to have disappeared in the late-time post-explosion images in all the
four bands where originally detected ($I$, $J$, $H$, and $K_{\rm s}$). In
these bands, the SN had already faded-away whereas it was still marginally
detected in the late-time $V$ and $R$ band images. To better visualize the
data all the images were scaled to a common pixel scale of 0.20'' pixel$^{-1}$
and the pre-explosion $K_{\rm s}$ image with the best seeing was convolved to
match the point spread functions (PSF) of the pre-explosion $I$-band
observation using the ISIS image subtraction software (S6). Both the pre-
and post-explosion $V$, $I$, and $K_{\rm s}$-band images were color-combined
using the DS9 tool\footnote{http://hea-www.harvard.edu/RD/ds9/} and are shown in
{\it Fig. 1}.

\begin{quote}
{\bf References and Notes}\\
\smallskip
S1. L.A.G. Monard, {\it CBET} {\bf 1315}, 1 (2008).\\
S2. N. Morrell, M. Stritzinger, {\it CBET} {\bf 1335}, 1 (2008).\\
S3. W. Li et al., {\it CBET} {\bf 1319}, 1 (2008).\\
S4. D. Maoz, F. Mannucci, F., {\it ATel} {\bf 1464} (2008).\\
S5. S. Mattila et al., {\it ApJ} {\bf 688L}, 91 (2008).\\
S6. C. Alard., R.H. Lupton. {\it ApJ} {\bf 503}, 325 (1998).\\
\end{quote}

\bibliography{scibib}

\bibliographystyle{Science}

\clearpage

\begin{table}
\caption{Pre- and post-explosion observations of the site of SN 2008bk.}
\begin{tabular}{llllll}
\hline\hline
Date (UT) & Telescope/ & Filter  & Exp. Time & FWHM     & resolution \\
          & Instrument &         & (sec)     & (arcsec) & (arcsec/pixel)\\
\hline
\multicolumn{4}{c}{{\bf Pre-explosion}} \\
2001 Sept 16.0 & VLT/FORS1   & $B$       & 300       & 1.2 & 0.20\\
2001 Sept 16.0 & VLT/FORS1   & $V$       & 300       & 1.0 & 0.20\\
2001 Sept 16.0 & VLT/FORS1   & $I$       & 480       & 0.9 & 0.20\\
2005 Apr 21.6  & VLT/ISAAC   & $J$       & 17$\times$60 & 0.5 & 0.148\\
2005 Oct 17.1  & VLT/ISAAC   & $K_{\rm s}$ & 58$\times$60 & 0.4 & 0.148\\
2007 Oct 16.1  & VLT/HAWKI   & $H$       & 2$\times$60  & 0.8 & 0.1064\\\hline
\multicolumn{4}{c}{{\bf Post-explosion}} \\
2008 May 19.4 & VLT/NACO     & $K_{\rm s}$ &  20$\times$60 & 0.1 & 0.027\\
\hline
\multicolumn{4}{c}{{\bf Late time post-explosion}} \\
2010 Sept 16.2 & NTT/EFOSC2  & $V$      & 8$\times$120 & 0.8 & 0.24\\
2010 Sept 16.2 & NTT/EFOSC2  & $R$      & 5$\times$120 & 0.9 & 0.24\\
2010 Sept 16.2 & NTT/EFOSC2  & $I$      & 4$\times$120 & 1.1 & 0.24\\
2010 Oct 29.0  & NTT/SOFI    & $J$      & 29$\times$60 & 1.5 & 0.288\\
2010 Oct 29.1  & NTT/SOFI    & $H$      & 30$\times$60 & 1.2 & 0.288\\
2010 Oct 29.1  & NTT/SOFI    & $K_{\rm s}$& 30$\times$60 & 1.2 & 0.288\\
\hline\hline
\end{tabular}
\end{table}